%% file: article.tex
\documentclass[aps,preprintnumbers,superscriptaddress,longbibliography,amsmath,amssymb,twocolumn,10pt,floatfix]{revtex4-1}

\usepackage{graphicx}
\usepackage{dcolumn}
\usepackage{bm}
\usepackage{xspace}
\usepackage{multirow}
\usepackage{booktabs}
\usepackage{xcolor}
\usepackage{xfrac}

\usepackage{comment}
\usepackage{physics}
\usepackage{xcolor}

\usepackage[separate-uncertainty = true]{siunitx}
\DeclareSIUnit{\depth}{\gram\per\square\centi\meter}

\usepackage[mathlines]{lineno}

\newcommand{\alphahad}{\alpha_{\text{had}}}
\newcommand{\alphaem}{\alpha_{\text{EM}}}
\newcommand{\zetahad}{\zeta_{\text{had}}}
\newcommand{\zetaem}{\zeta_{\text{EM}}}
\newcommand{\multhad}{m_{\text{had}}}
\newcommand{\multem}{m_{\text{EM}}}
\newcommand{\multtotal}{m_{\text{total}}}
\newcommand{\elasticity}{\kappa_{\text{el}}}
\newcommand{\etahad}{\eta_{\text{had}}}
\newcommand{\etaem}{\eta_{\text{EM}}}

\newcommand{\conex}{\textsc{Conex}}
\newcommand{\eposr}{E\textsc{pos\,}LHC-R\xspace}
\newcommand{\epos}{E\textsc{pos\,}LHC\xspace}
\newcommand{\qIII}{QGS\textsc{jet}-III.01\xspace}
\newcommand{\qII}{QGS\textsc{jet}-II.04\xspace}
\newcommand{\qIIthree}{QGS\textsc{jet}-II.03\xspace}
\newcommand{\qgsone}{QGS\textsc{jet}\,01\xspace}
\newcommand{\sibe}{S\textsc{ibyll}\,2.3e\xspace}
\newcommand{\sibd}{S\textsc{ibyll}\,2.3d\xspace}
\newcommand{\sibc}{S\textsc{ibyll}\,2.3c\xspace}
\newcommand{\sib}{S\textsc{ibyll}\,2.1\xspace}

\newcommand{\xmax}{X_{\max}}
\newcommand{\dxmax}{\Delta X_{\max}}

\newcommand{\x}[1]{%
  {}$
  \kern-2\mathsurround 
  $
  \binoppenalty10000 \relpenalty10000 #1
  {}$
  \kern-2\mathsurround 
  $
}

\usepackage{hyperref}
\hypersetup{
    colorlinks=true,
    linkcolor=blue,
    filecolor=blue,
    urlcolor=blue,
    pdftitle={Overleaf Example},
    pdfpagemode=FullScreen,
    }

\newcommand{\suppref}{\hyperref[apx:SM]{Supplemental Material}}

\begin{document}

\title{Proton–air interaction properties at {\boldmath $\sqrt{s} \simeq 100$} TeV from shower-depth measurements with the Pierre Auger Observatory and their connection to the Muon Puzzle}

\input{revtex_authorlist}

\date{\today}

\begin{abstract}
Hybrid measurements at the Pierre Auger Observatory indicate that most high-energy hadronic interaction models underestimate the average depth of the shower maximum, $\expval{\xmax}$, at a center-of-mass energy of $\sqrt{s}=97.7 \pm 0.4^{+6.6}_{-6.2}\,\mathrm{TeV}$. In this Letter, the hadronic interaction models are shown to follow a universal relation between the predicted $\expval{\xmax}$ and the mean values of variables characterizing the energy spectra of secondary particles produced in the first interaction of proton-induced air showers. Assuming the validity of these relations in Nature, we map the values of $\expval{\xmax}$ favored by Auger data into mean values of these variables. All models favor an increase in the mean elasticity and in the fraction of hadronic energy in proton--air interactions. The latter must be amplified by a factor of $2.8$ to $4.6$ to account for the muon puzzle.
\end{abstract}

\pacs{Valid PACS appear here}
\maketitle


\section{Introduction} \label{sec:intro}

Extensive air showers (EAS), initiated by ultra-high-energy cosmic rays, are governed by the production and reinteraction of hadrons in regions of phase space poorly constrained by accelerator data. As a result, extrapolations of high-energy hadronic interaction models used to interpret EAS measurements introduce significant uncertainties and tensions in their description~\cite{2022_Albrecht_MuonPuzzle, 2024_Auger_Xmaxs1000fits}.
\par
Two key air-shower observables sensitive to multiparticle production at
extreme energies are the depth of shower maximum $\xmax$ and the number of muons at ground $N_\mu$. Data from the Pierre Auger Observatory~\cite{2015_Auger_PAODescription} show evidence for a $>5\sigma$ discrepancy between the measured joint distribution of $\xmax$ and ground signal and the predictions of recent hadronic interaction models in the energy range $E_0 \in [10^{18.5},\,10^{19.0}]\,\mathrm{eV}$~\cite{2025_Auger_UpdateXmaxShifts}. Agreement with data can be achieved by deepening $\expval{\xmax}$ and increasing the predicted hadronic ground signal, implied by a larger $\expval{N_\mu}$~\cite{2024_Auger_Xmaxs1000fits}. This solution assumes a primary mass composition bounded by proton and iron, and shower-to-shower fluctuations fixed to the predictions of each hadronic interaction model. There is no evidence of a primary-mass-dependence of the required shifts~\cite{2025_Auger_UpdateXmaxShifts}, and they partially alleviate the long-standing \textit{Muon Puzzle}~\cite{2022_Albrecht_MuonPuzzle}.
\par
This Letter interprets the $\expval{\xmax}$ shifts using
universal relations, observed across high-energy hadronic interaction
models, between $\expval{\xmax}$ and mean production variables of the first proton-air interaction. These relations arise because the
models, despite different implementations, share basic conservation
laws and physical mechanisms, and constraints from accelerator data.
Assuming that these relations are present in Nature, we use them to map the values of $\expval{\xmax}$ favored by Auger data into modified proton--air
production variables. This interpretation assumes that the proton--air
cross section is correctly described by models~\cite{2012_Auger_xsection, 2023_Olena_xsection_comp}
and that the properties of the first interaction propagate continuously to
lower energies in subsequent cascade interactions. These assumptions
could fail if $\expval{\xmax}$ were to change appreciably without a corresponding modification of the
first proton--air interaction due to exotic physics, or sizeable modifications confined to later hadron--air interactions such as pion--air interactions. The inferred modifications provide a
consistent explanation of Auger data in terms of enhanced far-forward
hadronic energy flow at $\sqrt{s}\simeq \SI{100}{TeV}$ and can be
partially tested, at lower energies and in related collision systems, by
forward and far-forward accelerator measurements.

\par
This paper is organized as follows. Sec.~\ref{sec:first_interaction} introduces \textit{production variables} that characterize the energy spectrum of secondary particles in the first proton--air interaction. Sec.~\ref{sec:simulations} establishes universal relations between their expected values and $\expval{\xmax}$ using air-shower simulations. Sec.~\ref{sec:application} presents the Pierre Auger Observatory and applies the relations derived in Sec.~\ref{sec:simulations} to interpret the shifts in $\expval{\xmax}$ preferred by data in terms of modifications of particle production in the primary interaction of each hadronic interaction model. The resulting increase in the hadronic energy fraction is then used to estimate the enhancement of energy in the hadronic shower needed to reproduce the observed muon content. Sec.~\ref{sec:conclusions} summarizes our main findings.

\section{First interaction variables}
\label{sec:first_interaction}
The first proton--air interaction produces $\multtotal$ secondary particles, of which $\multhad$ tend to sustain the hadronic cascade through re-interactions, while $\multem$ feed the electromagnetic (EM) cascade. The EM sector comprises $\pi^0$, $\eta$, and their dominant decay products, $\gamma$ and $e^\pm$. Notably, neutral pions carry $\sim 90\%$ of the energy transferred to the EM cascade.
\par
Using the fractions of the primary energy carried by particles in each sector in the lab. frame, we define the fraction of energy carried by hadronically interacting secondaries $\alphahad \equiv \sum_{i=1}^{\multhad} x_i$~\cite{2018_Cazon_alpha}, the elasticity $\elasticity$ as the energy fraction of the leading secondary particle, and
\begin{equation} \label{eq:zeta_definitions}
    \zetahad \equiv -\sum_{i=1}^{\multhad} x_i \ln x_i \quad,\quad
    \zetaem \equiv -\sum_{j=1}^{\multem} x_j \ln x_j,
\end{equation}
where $x_i$ ($x_j$) denotes the fraction of the primary energy carried in the lab. frame by the $i$-th hadronic ($j$-th EM) particle.
\par
The connection between $\alphahad$ and the muon content of showers, including the role of subsequent shower generations, is discussed in Sec.~\ref{sec:application}. The variables $\zetahad$ and $\zetaem$ encode the stochastic energy partition among secondary particles in the primary proton--air interaction in a way that maximally correlates with the shower-to-shower values of $\xmax$~\cite{2025_miguel_xmaxmodel}. Their functional forms are motivated by the Heitler--Matthews framework~\cite{2005_Matthews_HMmodel}, with each term $x \ln x$ representing the energy-weighted contribution of a secondary particle to $\xmax$. These definitions resemble the Shannon entropy~\cite{1948_Shannon_entropy}, but use random variables $x_i$ rather than their probabilities. This analogy facilitates the derivation and interpretation of the properties of the $\zeta$ variables~\cite{2025_miguel_xmaxmodel}. Lastly, these production variables can be measured in the kinematic phase space accessible to particle accelerators.

\section{Dependence of \boldmath{$\expval{\xmax}$} on first interaction variables}
\label{sec:simulations}
Using LHC-tuned high-energy hadronic interaction models, we study the dependence of $\expval{\xmax}$ on the mean values of the production variables of primary proton--air interactions introduced in Sec.~\ref{sec:first_interaction}. These models span a physically motivated parameter space, constrained by conservation laws, internal consistency, and agreement with accelerator data in the appropriate kinematic limits. Thus, our study accounts for plausible correlations between production variables, complementing the approaches of Refs.~\cite{2011_Ulrich_f19, 2023_Jan_Mochi, 2024_Riehn_sibyllstar, 2026_Blazek_mochi}, which probe a broader parameter space by freely varying interaction parameters, including more extreme configurations.
\par
To this end, we generated $10^4$ proton-induced showers with \conex{} v7.50 and v7.80~\cite{2007_Bergmann_conex,2004_Pierog_conex} (according to the hadronic interaction model), with primary energy $E_0 = 10^{18.7}$~eV and zenith angle $\theta = 55^\circ$, using the hadronic interaction models: \eposr{}~\cite{2025_Tanguy_eposlhcr}, \qIII{}~\cite{2024_Ostapchenko_qgsIII}, and \sibe{}~\cite{2020_Felix_sibyll23d}\footnote{This version handles hadronic interactions in the same manner as \sibd{}, but it corrects a feature affecting the particle signal at ground. This could lead to different shifts in $\expval{\xmax}$ required of both models, so both models were kept in this Letter.} — and their earlier versions — \epos{}~\cite{2015_Pierog_eposlhc}, \qII{}~\cite{2011_Ostapchenko_qgsjet}, and \sibd{}~\cite{2020_Felix_sibyll23d}. The chosen primary energy and zenith angle correspond, respectively, to the average reconstructed energy of the data used in Ref.~\cite{2024_Auger_Xmaxs1000fits}: $\log_{10}(\expval{E_{\rm{rec}}} / \unit{\eV}) = 18.706\pm 0.003 \pm 0.057$, corresponding to the nucleon-nucleon center-of-mass energy $\sqrt{s} = 97.7 \pm 0.4_{-6.2}^{+6.6} \unit{\TeV}$, and to the center of the largest zenith-angle bin of Ref.~\cite{2024_Auger_Xmaxs1000fits}. The high zenith angle ensures the development of the muonic cascade after its maximum, minimizing truncation effects by the ground.
\par
\begin{figure*}[ht!]
    \centering
    \includegraphics[width=\linewidth]{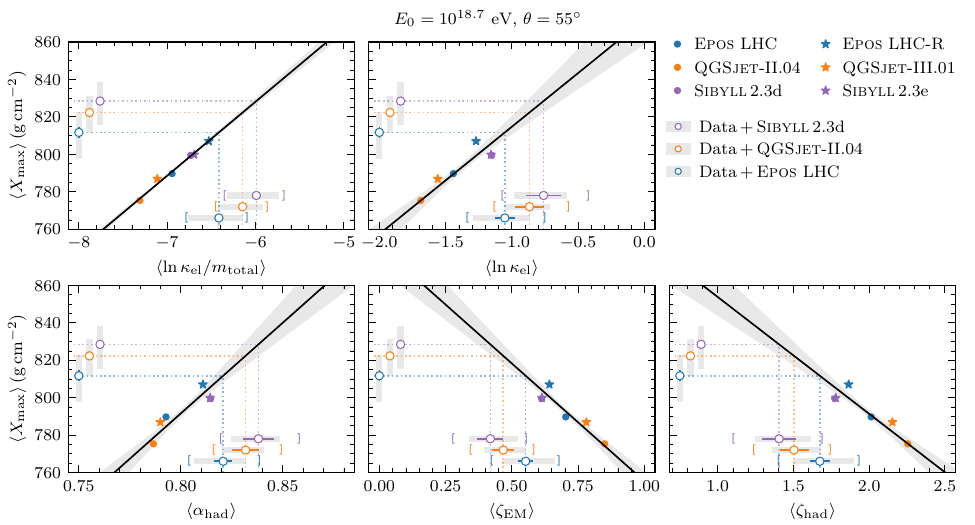}
    \caption{ Values of $\expval{\xmax}$ against the multiparticle production variables $\expval{\ln \elasticity / \multtotal}$ (top left), $\expval{\ln \elasticity}$ (top middle), $\expval{\alphahad}$ (bottom left), $\expval{\zetaem}$ (bottom middle) and $\expval{\zetahad}$ (bottom right) for different high-energy hadronic interaction models. The most recent model versions are represented by the starred markers, and their earlier versions by the circular markers. The fit curve is represented in black, and its uncertainty is represented by the grey band. Empty markers (``Data") are the shifted values of $\expval{\xmax}$ determined in~\cite{2024_Auger_Xmaxs1000fits} for \epos{}, \qII{} and \sibd{} together with their projection onto the $x$-axis along the fit curves. Highly correlated systematic uncertainties are represented by shaded grey bars. The increase of these systematic uncertainties caused by the uncertainty of the fit curve is represented by the colored brackets.}
    \label{fig:new_primary_vars_xmax_calibration}
\end{figure*}

Fig.~\ref{fig:new_primary_vars_xmax_calibration} shows the dependence of $\expval{\xmax}$ on $\expval{\ln (\elasticity / \multtotal)}$ (following Ref.~\cite{2012_Kampert_MassModelDep}), $\expval{\ln \elasticity}$, $\expval{\alphahad}$, $\expval{\zetahad}$, and $\expval{\zetaem}$. The statistical uncertainty in the predicted mean values of the production variables was neglected, as it remains below $1\%$ owing to the large number of simulated showers. The details of the fitting procedure and the linear relations between $\expval{\xmax}$ and additional production variables are provided in the \suppref{}.
\par
\par
The universality of the relation between $\expval{\xmax}$ and the mean of the production variables of the first interaction is quantified by the standard deviation of the residuals with respect to the fitted curves. Depending on the production variable, these standard deviations range from $\SI{1.7}{\depth}$ to $\SI{5.2}{\depth}$ and are partially driven by the spread in model predictions of the proton--air cross section. This spread is largely independent of the different implementations of hadron production in the primary interaction. In all cases, the dispersion of
$\expval{\xmax}$ about the lines is much smaller than the
range covered by the hadronic interaction models,
$\sim \SI{34}{\depth}$, and smaller than the systematic uncertainty in the
measurement of $\xmax$, $\sim \SI{8}{\depth}$~\cite{2014_Auger_xmax}.
This small dispersion reflects the strong universality of the relations: the
effect of later shower generations on $\expval{\xmax}$ is proportional to the effect of the first interaction, with a proportionality constant nearly independent of the hadronic interaction model. Thus, the explicit dependence on the latter can be replaced by a dependence on first-interaction properties. Lastly, the quantity $\expval{\ln (\elasticity / \multtotal)}$ yields the smallest dispersion, while $\expval{\ln \elasticity}$ gives the largest. By contrast, the relation between $\expval{\xmax}$ and $\expval{\ln \multtotal}$ is strongly model dependent, precluding a clear connection between the two quantities. The figures supporting this claim and the numerical values provided throughout the paragraph are included in the \suppref{}.
\par
The universality of the discovered relations is supported by the continuous energy evolution of subsequent
interactions down to kinematic regimes where model predictions converge to accelerator data, physical mechanisms common to all models, and the rapid decoupling of the
electromagnetic cascade from the hadronic core~
\cite{2016_Ostapchenko_XmaxScale_1stInt}, which prevents residual model
differences from being amplified along the shower. In contrast, additional
standard-physics modifications, including changes in pion--air interaction cross
sections or particle production, would increase the dispersion around the curves shown in Fig.~\ref{fig:new_primary_vars_xmax_calibration}.
Because pion--air interactions are mainly constrained by low-energy
fixed-target experiments~\cite{2014_NA61_design, 2017_NA61_piCresonances} and are crucial for muon production, they
remain an important source of freedom in hadronic interaction models. Differences in the pion--air interaction cross-section might already contribute to the residual dispersion observed around the curves. Within the freedom set by the collection of current models, the impact of changing this parameter on $\expval{\xmax}$ is conservatively estimated to be $\sim \SI{10}{\depth}$ (see \suppref{}).
\par
Lastly, the universal correspondence between $\expval{\xmax}$ and multiparticle-production variables of the first interaction was further validated in the \suppref{} using pre-LHC hadronic interaction models. The inclusion of legacy models produces only a mild increase in the dispersion around the curves, consistent with the absence of LHC constraints and known shortcomings of earlier model generations, rather than with differences in the proton--air cross section.
\section{Application to data}
\label{sec:application}
Hybrid data from the Pierre Auger Observatory favor a significant shift in $\expval{\xmax}$ and a rescaling of the muon content predicted by air-shower simulations to reproduce the joint distribution of $\xmax$ and ground particle signals~\cite{2024_Auger_Xmaxs1000fits, 2025_Auger_UpdateXmaxShifts}. These results are based on 2239 hybrid events recorded between 1 January 2004 and 31 December 2018, with energies $E_0 \in [10^{18.5}, 10^{19.0}]$~eV and zenith angles $\theta \in (0^\circ, 60^\circ)$. The reconstructed values of $\xmax$ and primary energy are provided by the Fluorescence Detector (FD)~\cite{2010_Auger_FDdescription} at four sites of the observatory. These events are simultaneously recorded by the Surface Detector array (SD-1500)~\cite{2008_Auger_SDdescription}, comprising 1\,600 water-Cherenkov detectors arranged on a triangular grid with $\SI{1.5}{\km}$ spacing and covering $\SI{3000}{\km^2}$. Data selection criteria and corrections are described in Ref.~\cite{2024_Auger_Xmaxs1000fits}, which details the main assumptions of the analysis: a primary composition consisting of proton, helium, oxygen, and iron nuclei, and fixed shower-to-shower fluctuations in $\xmax$ and the ground signal during the fitting procedure for each hadronic interaction model.
\par

Since Refs.~\cite{2024_Auger_Xmaxs1000fits, 2025_Auger_UpdateXmaxShifts} show no evidence that the observed $\xmax$ shifts depend on the primary mass composition, and measurements of the proton--air cross section are compatible with model predictions~\cite{2012_Auger_xsection, 2023_Olena_xsection_comp}, we use the framework developed in Sec.~\ref{sec:simulations} to quantify the contribution of the first proton--air interaction to the observed shifts in $\expval{\xmax}$. This interpretation assumes that modifications to hadronic interactions evolve continuously with projectile energy throughout the shower cascade. The assumption could fail in the presence of exotic physics or modifications affecting later hadron--air interactions while leaving proton--air interactions unchanged. For example, changes to pion--air interactions can affect mildly the overall scale of $\xmax$ without affecting the first proton--air interaction.

\par
Extrapolating the linear relations shown in Fig.~\ref{fig:new_primary_vars_xmax_calibration} yields the values of $\expval{\ln (\elasticity / \multtotal)}$, $\expval{\ln \elasticity}$, $\expval{\alphahad}$, $\expval{\zetahad}$, and $\expval{\zetaem}$ corresponding to the measured shifts in $\expval{\xmax}$. These values are indicated by the open colored markers for \epos{}, \qII{}, and \sibd{} . The colored error bars represent the statistical uncertainties of the shifted $\expval{\xmax}$ values and of the inferred production variables, including the propagated uncertainty of the curves.
\par
\par
The systematic uncertainties in the $\expval{\xmax}$ shifts are strongly correlated among models because they are dominated by the common systematic uncertainties, in particular, the $14\%$ systematic uncertainty in the FD energy scale~\cite{2020_Auger_sdSpectrum}. These common contributions are represented by the gray bars, while the colored brackets indicate their enlargement due to the uncertainty of the fitted lines. These lines themselves are affected by the uncertainty of the FD energy scale~\cite{2020_Auger_sdSpectrum}, since the relations between $\expval{\xmax}$ and the mean production variables form a family of nearly parallel and universal lines in the energy range $E_0 \in [10^{17},\,10^{19}]$~eV. As discussed in the \suppref{}, the resulting contribution to the systematic uncertainty of the inferred production variables is at most $\sim6\%$ and is therefore neglected in the final values.
\par
The deeper values of $\expval{\xmax}$ preferred by the data suggest increased values of $\expval{\ln(\elasticity/\multtotal)}$, $\expval{\ln\elasticity}$, and $\expval{\alphahad}$, and lower values of $\expval{\zetaem}$ and $\expval{\zetahad}$, irrespective of the hadronic interaction model. The final mean values of the production variables are incompatible with the default model predictions, given the incompatibility between the shifted and nominal values of $\expval{\xmax}$~\cite{2024_Auger_Xmaxs1000fits} and the small uncertainty of the curves. The updated version \qIII{} fails to meet the modifications required for \qII{}, while \eposr{} is compatible with the modifications required for \epos{}. Indeed, \eposr{} has a higher frequency of diffractive hadron-air interactions and a smaller pion-air cross-section compared to all the other models, both needed to reproduce the $\xmax$-scale preferred by Auger data~\cite{2025_Tanguy_eposlhcr}. Particle production in \sibe{} is implemented exactly as in \sibd{}, so both models remain incompatible with data.
\par
These results indicate that increasing the fraction of energy transferred to the hadronic sector of the primary interaction is not sufficient to explain the results of Ref.~\cite{2024_Auger_Xmaxs1000fits}. Rather, the energy spectra of secondary particles within each sector must also be modified to accommodate the predicted decrease in $\expval{\zetahad}$ and $\expval{\zetaem}$. The data favor a more asymmetric partition of the primary energy, in which a small number of particles carry a large fraction of the available energy while the remaining secondaries are shifted toward very low energies. Such configurations yield low $\zeta$ values since $-x\ln x \to 0$ for both $x\to0$ and $x\to1$, and are associated with diffractive interactions, rapidity gaps, or strongly asymmetric non-diffractive collisions.
\par
Equivalently, the preference for low $\expval{\zetahad}$ and high $\expval{\alphahad}$ values implies an enhanced hadronic energy flow toward larger pseudorapidities, accompanied by the production of very-low-energy hadrons at central rapidities. This stems from the relation $\zetahad \simeq - \eta_{\rm had} + \alphahad \ln\left(2E_0 /Q\right)$,
where $\eta_{\rm had}$ is the total energy-weighted hadronic pseudorapidity
and $Q \approx \SI{250}{\MeV}$ is a transverse momentum scale~\cite{2025_miguel_xmaxmodel} and is explored in the
\suppref{}. Such energy flows could be probed experimentally, albeit at lower energies, in forward and far-forward collider experiments such as FASER~\cite{2024_FASER_design, 2025_FASER_muonNeutrinoFlux} and LHCf~\cite{2008_LHCf_design, 2016_LHCf_forward_npions, 2020_LHCf_forwardneutrons}, in particular in proton--oxygen runs at the LHC. In fact, data from these collisions have already enabled ATLAS measurements of the proton-oxygen interaction cross-section and particle production at central rapidity~\cite{2026_Atlas_pOxsection}, and will also be analysed by LHCf in the very forward region. Lastly, specific modifications to multiparticle production and their impact on production variables are left for future work.
\par
A better description of the data is achieved if the models \epos{}, \qII{}, and \sibd{} increase $\expval{\alphahad}$ by $3.6\%$, $6.0\%$, and $3.1\%$, respectively. These values are consistent with the reasoning of Ref.~\cite{2021_Auger_muonfluctuations}: the agreement between the predicted and measured fluctuations of $N_\mu$ suggests a small increase in $\alphahad$ over several air-shower generations, rather than a large increase in the first interaction. The amplification of a small change in $\expval{\alphahad}$ follows from
\begin{equation} \label{eq:muon_scale_alpha}
    {N_\mu} = \frac{E_0}{\xi_c^\pi}
    \prod_{g=1}^c\alpha_{\text{had},g},
\end{equation}
\noindent
where $g$ denotes the air-shower generation, $\alpha_{\text{had},g}$ the value of $\alphahad$ for all outgoing particles in generation $g$, and $c$ the critical generation at which muons are produced at the critical energy $\xi_c^\pi \sim \mathcal{O}\left(\SI{100}{\GeV}\right)$. Eq.~\eqref{eq:muon_scale_alpha} shows that $\expval{N_\mu}$ is determined by the energy retained in the hadronic core of the shower across all generations $g = 1,2,\dots,c$, i.e., by $\alpha_{\text{had},g}$.
\par
Using Eq.~\eqref{eq:muon_scale_alpha}, the increase in $N_\mu$, $\delta \ln N_{\mu}$, can be related to $\delta \ln \expval{\alphahad}$ by assuming $\ln \expval{\alphahad} \to \ln \expval{\alphahad} + \delta \ln \expval{\alphahad}$ and $\ln \alpha_{\text{had}, g} \to \ln \alpha_{\text{had}, g} + f_g \times \delta \ln \expval{\alphahad}$, where $f_g$ denotes the modification factor for generation $g$, taking the value $f_1=1$ and decreasing progressively with increasing $g$. This yields
\begin{equation}
\delta \ln \expval{N_\mu} = g_{\text{mod}} \times \delta \ln \expval{\alphahad}
\label{eq:muonscaling}
\end{equation}
where $g_{\text{mod}}=\sum f_g$ is the sum of the modification factors over all shower generations $g$.
Thus, $g_{\rm mod}$ is an amplification factor that maps a shift in
$\ln \expval{\alphahad}$ into a rescaling of the muon number,
$\expval{N_\mu} \rightarrow R_\mu \expval{N_\mu}$, with $R_\mu = \exp\!\left( g_{\text{mod}} \times \delta \ln \expval{\alphahad} \right)$.
\par
Using Eq.~\eqref{eq:muonscaling}, we estimate the values of $g_{\rm mod}$ that match the muon rescalings inferred by Auger:
$R_{\mu}=15\pm1_{-10}^{+13}\%$, $17\pm1_{-11}^{+14}\%$, and $14\pm1_{-10}^{+14}\%$ for \epos{},
\qII{}, and \sibd{}, respectively. These values were obtained from the hadronic rescaling reported in Ref.~\cite{2024_Auger_Xmaxs1000fits} (see the \suppref{}).
\par
\begin{figure}[t!]
    \centering
    \includegraphics[width=\linewidth]{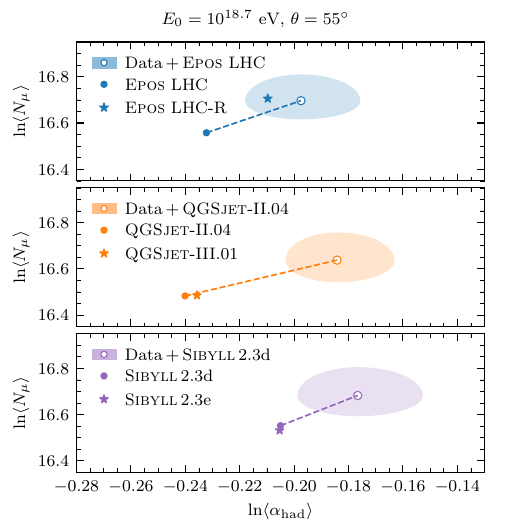}
    \caption{ $\ln \expval{\alphahad}$ vs $\ln \expval{N_\mu}$ for several hadronic interaction models. The solid markers correspond to the default values for each hadronic interaction model and the empty markers to the shifted values determined in Ref.~\cite{2024_Auger_Xmaxs1000fits} and this work. The slope of the dashed lines gives the amplification of the increase in $\ln \expval{\alphahad}$ required to account for the measured $\ln \expval{N_\mu}$.}
    \label{fig:alpha_nmu_calibration_per_model}
\end{figure}
\par
Fig.~\ref{fig:alpha_nmu_calibration_per_model} displays the nominal and modified values of $\ln \expval{\alphahad}$ vs $\ln \expval{N_\mu}$, including the $1\sigma$ uncertainty of the latter, and the straight line connecting them. The slope of this line is the amplification factor and reads $g_{\text{mod}} = \{4.0_{-2.5}^{+7.5},\, 2.8_{-1.8}^{+2.9},\, 4.6_{-3.3}^{+18.6}\}$ for \epos{}, \qII{}, and \sibd{}, respectively. The large systematic uncertainties, particularly the upper ones, are driven by the uncertainties in $R_{\mu}$ and $\langle \alphahad\rangle$ (propagated from those on $\expval{\xmax}$) (see \suppref{}). The values of $g_{\text{mod}}$ are compatible between hadronic interaction models. \eposr{} satisfies the shifts in both $\expval{\alphahad}$ and $\expval{N_\mu}$ inferred for \epos{} for $\theta = 55^\circ$, but fails to reproduce the preferred muon-rescaling factor for more vertical showers~\cite{2025_Auger_UpdateXmaxShifts}.
\par
Within the accuracy set by the dispersion of the calibration curves in
Fig.~\ref{fig:new_primary_vars_xmax_calibration}, and assuming that the models capture
the relevant standard-physics mechanisms, the Auger
$\expval{\xmax}$ shifts translate into modifications of the
secondary-energy spectra in the primary interaction. Conceivable changes in the proton-air, pion–air and kaon-air interaction cross-sections in Nature would shift $\expval{\xmax}$ without changing, to first order, the values of the multiparticle production variables of the first proton-air interaction. This would correspond to an overall upward or downward shift in the linear relations of Fig.~\ref{fig:new_primary_vars_xmax_calibration}, in case the cross-section values decrease or increase, respectively. In the first case, the derived values of production variables should be viewed as an upper bound of the modification, and in the latter, to a lower one. These scenarios could be ruled out by considering their impact on $N_\mu$ or the depth of the maximum of the muon production depth distribution~\cite{2014_Auger_XmuMax, 2024_Ostapchenko_HIMuncertaintiesXmax}.
\par
Lastly, the derived shifts rely on a primary mass composition bounded by proton and iron. Thus, alternative interpretations of the primary mass composition~\cite{2025_Vicha_HeavyMetal, 2024_Farrar_BNSM, 2025_Farrar_BNSM} could modify these shifts and, consequently, the values of the production variables of the primary interaction. Nevertheless, the universality of the mapping between $\expval{\xmax}$ and the mean values of these variables would be preserved, ensuring the robustness of the framework presented here.

\section{Conclusions}
\label{sec:conclusions}
We demonstrate a universal relation between the average depth of shower maximum, $\expval{\xmax}$, and key variables characterizing the first proton--air interaction. This universality is robust and reflects the shared underlying physics of hadronic interaction models, which, despite differences in implementation (e.g., diffraction, hadronization, and energy partitioning), collectively span a physically plausible phase space.


Using these relations, the shifted values of $\expval{\xmax}$ reported by the Pierre Auger Collaboration were mapped into multiparticle production variables of the primary interaction. In the absence of new physics or large changes in pion--air interaction cross sections and multiparticle production, the observed shifts in $\expval{\xmax}$ can be reproduced through lower values of $\expval{\zetahad}$ and $\expval{\zetaem}$, together with higher values of $\expval{\ln(\elasticity / \multtotal)}$, $\expval{\ln \elasticity}$, and the hadronic energy fraction $\expval{\alphahad}$. These trends are common to all hadronic interaction models and consistently point towards enhanced secondary-particle production in the far-forward region of phase space, shifting the hadronic energy flow towards higher rapidities. The corresponding interactions distribute the projectile energy more asymmetrically among outgoing particles. Such effects can be directly probed by far-forward measurements in proton--oxygen collisions at the LHC. Amplified by $2.8$--$4.6$, the increase in $\expval{\alphahad}$ produces a surplus of hadronic energy sufficient to account for the Muon Puzzle.

Within the stated assumptions, the link between the observed shifts in $\expval{\xmax}$ and the energy distribution of secondary hadrons in the first proton--air interaction, is consistent with the Muon Puzzle and provides constraints on hadron production informed by air-shower data at $\sqrt{s} \simeq \SI{100}{\TeV}$.

\input{acknowledgments}

\bibliography{references}

\clearpage
\appendix
\onecolumngrid
\section*{Supplemental Material}\label{apx:SM}
\subsection{Fit algorithm}
The fit curve shown in Fig.~\ref{fig:new_primary_vars_xmax_calibration} of the main text, and all figures relating mean values of production variables of the primary interaction with $\expval{\xmax}$, is obtained as follows. First, we fit the model predictions to a linear function by minimizing a $\chi^2$-fit assuming a unit standard deviation as the error in $\expval{\xmax}$. The standard deviation of the residuals about this initial fit is then used to update the uncertainties in $\expval{\xmax}$, and the fit is repeated to determine the final curve and uncertainties of the fit parameters.
\par
This method is preferred to a simple $\chi^2$ fit using the statistical uncertainty of the predicted values of $\expval{\xmax}$ because the dispersion of the models about the fit curve does not come from statistical fluctuations. Rather, it stems from residual differences between hadronic interaction models.

\subsection{Compound multiparticle production variables} \label{apx:xi_k_xmax}
As detailed in Ref.~\cite{2025_miguel_xmaxmodel}, shower-to-shower fluctuations in the energy spectra of secondary particles produced in the primary interaction are captured by the estimator $\xi$ of the shower observable $\dxmax \equiv \xmax - X_1$, where $X_1$ is the depth of the first interaction. In this framework,
\begin{equation} \label{eq:xi_model_simple}
\begin{aligned}
\xi \simeq 917 - (311 - 9.6 \omega) \alphahad - 37 (\omega \zetahad + \zetaem)
\;\;[\unit{\depth}],
\end{aligned}
\end{equation}
\par\noindent
with $\omega = -0.140 \expval{\zetaem / (1 - \alphahad)} + 1.09$ for $E_0 = 10^{19}$~eV. The numerical coefficients are, in fact, slowly varying functions of the primary energy.
\par
\begin{figure}[!ht]
    \centering
    \includegraphics[width=5.5in]{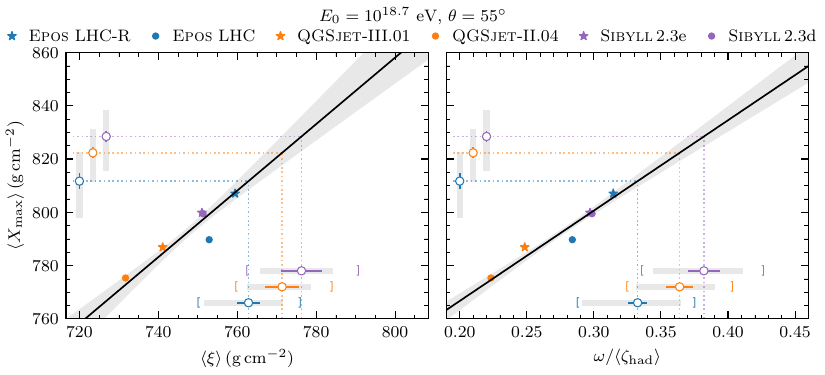}
       \caption{ Values of $\expval{\xmax}$ against $\expval{\xi}$ (left) and $\omega / \expval{\zetahad}$ (right) for different high-energy hadronic interaction models. The fit curve is represented in black, and its uncertainty, determined as explained in the main text, is represented by the grey band. Empty markers (``Data") are the shifted values of $\expval{\xmax}$ determined in~\cite{2024_Auger_Xmaxs1000fits} for \epos{}, \qII{} and \sibd{} together with their projection onto the $x$-axis. Highly correlated systematic uncertainties are represented by shaded grey bars. The increase of these systematic uncertainties caused by the uncertainty of the fit curve is represented by the colored brackets.}
    \label{fig:xmax_calibration_xi}
\end{figure}

Fig.~\ref{fig:xmax_calibration_xi} shows $\langle \xmax \rangle$ as a function of $\langle \xi \rangle$ (left) and $\omega / \expval{\zetahad}$ (right) for several hadronic interaction models.
\par
\par
Despite a small spread in the mean depth of the first interaction, $\langle X_1 \rangle$, of $\sim\SI{3}{\depth}$ at $E_0 = 10^{18.7}$ eV across all models, the standard deviations of the residuals of $\expval{\xmax}$ about the lines $\expval{\xi}-\expval{\xmax}$ and $\omega / \expval{\zetahad} - \expval{\xmax}$ are $\SI{5.1}{\depth}$ and $\SI{2.8}{\depth}$, respectively. This universality across models arises from common underlying physics in the hadronic models, with differences primarily driven by the primary interaction.
\subsection{The pion-air interaction cross-section}

\begin{figure}[th!]
    \centering
    \includegraphics[width=3.1in]{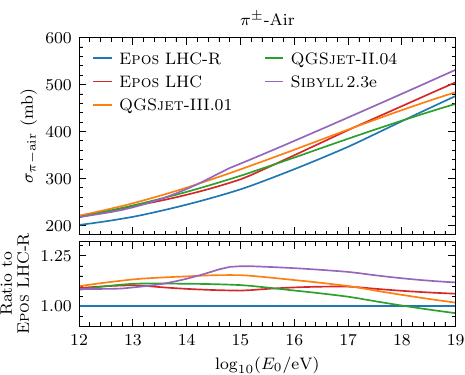}
    \caption{Pion-air interaction cross-section as a function of the projectile energy in the lab. frame, for several high-energy hadronic interaction models.}
    \label{fig:pionair_xsection}
\end{figure}

The energy dependence of the pion--air interaction cross section differs substantially among hadronic interaction models, as shown in Fig.~\ref{fig:pionair_xsection}. A decrease in $\sigma_{\pi\text{--air}}$ can deepen the overall scale of $\xmax$ through an increase in the interaction length of charged pions in air. However, despite the important role of pion--air interactions in air-shower development, the rapid decoupling of the electromagnetic component from the hadronic core suppresses the propagation of pion--air interaction differences into $\expval{\xmax}$~\cite{2016_Ostapchenko_XmaxScale_1stInt}.
\par
\par
In particular, despite the large differences in $\sigma_{\pi\text{--air}}$ predicted by \eposr{} and \sibe{}, their values of $\expval{\xmax}$ differ by only $\approx \SI{10}{\depth}$. The mean multiparticle production variables predicted by both models are nearly identical in the primary interaction and exhibit a similar energy evolution~\cite{2025_Miguel_PhDThesis}. This indicates that even sizeable variations in $\sigma_{\pi\text{--air}}$ produce only moderate deviations from the universal curves shown in Fig.~\ref{fig:new_primary_vars_xmax_calibration} of the main text.
\subsection{Legacy hadronic interaction models} \label{apx:old_models}
\begin{figure}[!ht]
    \centering
    \includegraphics[height= 2.25in]{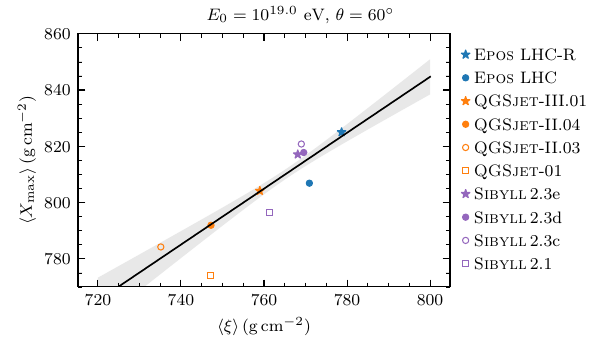}
    \caption{Values of $\expval{\xmax}$ against $\expval{\xi}$ for different high-energy LHC-tuned and pre-LHC hadronic interaction models. The fit curve is represented in black, and its uncertainty, determined as explained in the main text. The fit is performed only to the LHC-tuned hadronic interaction models \eposr{}, \epos{}, \qIII{}, \qII{}, \sibd{} and \sibe{}.}
    \label{fig:xmax_calibration_older_models_xi}
\end{figure}

\begin{figure*}[!ht]
    \centering
    \includegraphics[width= \linewidth]{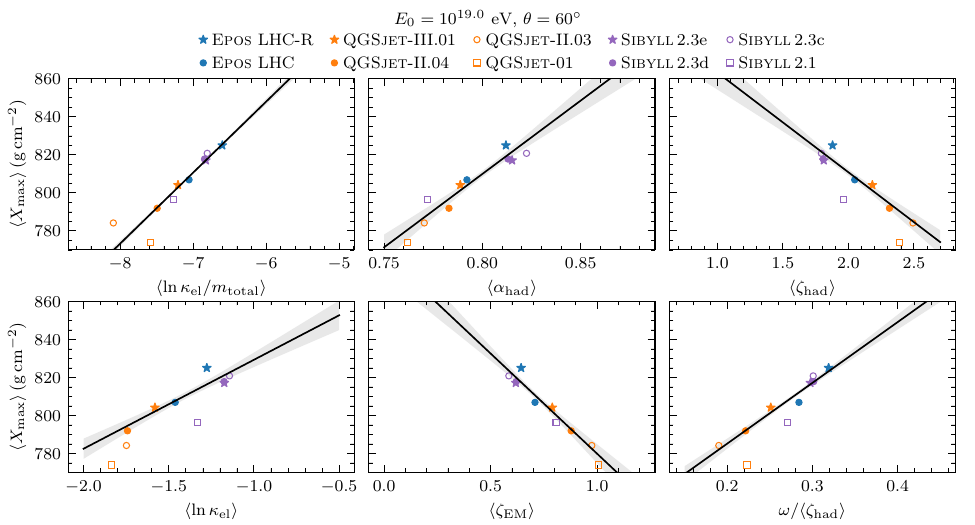}
    \caption{Values of $\expval{\xmax}$ against $\expval{\ln \elasticity / \multtotal}$ (top left), $\expval{\alphahad}$ (top center), $\expval{\zetahad}$ (top left), $\expval{\ln \elasticity}$ (bottom left), $\expval{\zetaem}$ (bottom center) and $\omega / \expval{\zetahad}$ (bottom right) for different high-energy LHC-tuned and pre-LHC hadronic interaction models. The fit curve is represented in black, and its uncertainty, determined as explained in the main text. The fit is performed only to the LHC-tuned hadronic interaction models \eposr{}, \epos{}, \qIII{}, \qII{}, \sibd{} and \sibe{}.}
    \label{fig:xmax_calibration_older_models_new_prodvars}
\end{figure*}

\begin{figure*}[!ht]
    \centering
    \includegraphics[width= \linewidth]{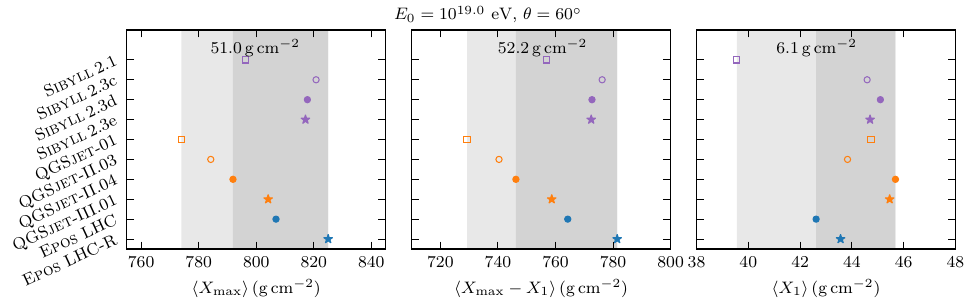}
    \caption{Values of $\expval{\xmax}$, $\expval{\dxmax}$ and $\expval{X_1}$ for different high-energy LHC-tuned and pre-LHC hadronic interaction models. The light grey band shows the spread in the mean values across all high-energy hadronic interaction models, while the darker grey regions correspond to the spread across the LHC-tuned hadronic interaction models used in the main text: \eposr{}, \epos{}, \qIII{}, \qII{}, \sibd{} and \sibe{}.}
    \label{fig:xmax_xfirst_spread_all_models}
\end{figure*}

The universality of the relations between $\expval{\xmax}$ and multiparticle-production variables of the first interaction is further validated including the pre-LHC models \sib{}~\cite{2009_Ahn_sibyll21}, \qgsone{}~\cite{1993_Ostapchenko_qgsjet01_theo, 1997_Ostapchenko_qgsjet01_implementation}, and \qIIthree{}~\cite{2006_Ostapchenko_qgsII03}, together with the LHC-tuned model \sibc{}~\cite{2019_Anatoli_sibyll23c}. The corresponding bands are shown in Figures~\ref{fig:xmax_calibration_older_models_xi} and \ref{fig:xmax_calibration_older_models_new_prodvars}. The fit curves are obtained using only the hadronic interaction models considered in the main text, while the legacy models provide an independent validation of the universality claim. The \conex{} air-shower simulation libraries were generated at $E_0 = 10^{19}$~eV and $\theta = 60^\circ$, according to the availability of simulations using deprecated hadronic interaction models. For completeness, the spread in the values of $\expval{\xmax}$, $\expval{\dxmax}$, and $\expval{X_1}$ is shown in Fig.~\ref{fig:xmax_xfirst_spread_all_models}.
\par
Including legacy models produces only a modest increase in the dispersion around the curves, consistent with the absence of LHC constraints and the known shortcomings of earlier model generations. The increased spread in the proton--air cross section does not explain this effect. In particular, the lower values of $\expval{\xmax}$ predicted by \sib{} can be partially attributed to its large proton--air interaction cross section~\cite{2020_Felix_sibyll23d}, whereas the remaining dispersion is instead associated with differences in particle production that are disfavored by LHC data. The mild violation of the universality of the curves nevertheless indicates that they originate from basic physical principles. Moreover, more recent models generally predict deeper values of $\expval{\xmax}$, bringing them closer to Auger data.
\par
\par
\par
\par
\subsection{Relation between {\boldmath $\expval{\xmax}$} and mean multiplicity in the primary interaction}
\label{apx:xmax_mult}
\begin{figure}[!ht]
    \centering
    \includegraphics[height= 2.25in]{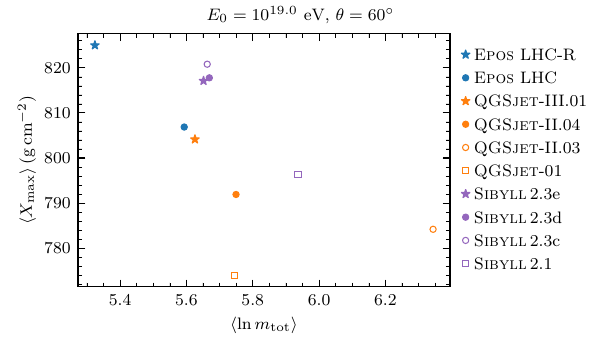}
    \caption{ $\expval{\ln \multtotal}$ vs $\expval{\xmax}$ for different high-energy LHC-tuned and pre-LHC hadronic interaction models.}
    \label{fig:xmax_calibration_older_models_mult}
\end{figure}

Fig.~\ref{fig:xmax_calibration_older_models_mult} shows a scatter plot of $\expval{\xmax}$ versus $\expval{\ln \multtotal}$, illustrating the weak correlation between these quantities.

\subsection{Propagation of the FD energy-scale systematic uncertainty}

\begin{figure}[ht!]
    \centering
    \includegraphics[width=3.4in]{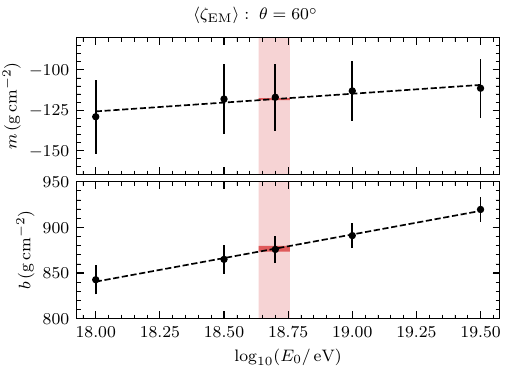}
    \caption{Primary-energy dependence of the slope, $m$, and intercept, $b$, of the relation between $\expval{\zetaem}$ and $\expval{\xmax}$. The dashed lines show linear fits to $m$ and $b$ as functions of $\log_{10}(E_0/\mathrm{eV})$. The light red band indicates the FD energy-scale uncertainty, while the darker red region shows the corresponding propagated uncertainty in $m$ and $b$.}
    \label{fig:calib_params_energy_func_zetaem}
\end{figure}

The universality of the relations between $\expval{\xmax}$ and the mean values of the production variables of the first proton--air interaction, $\expval{v}$, used in the main text remain valid, at least, in the primary energy range $E_0 \in [10^{17.0}, 10^{19.5}]~\mathrm{eV}$. However, the slope, $m$, and intercept, $b$, of the relation, $\expval{\xmax} = m \expval{v} + b$, exhibit a mild dependence on energy. This is illustrated in Fig.~\ref{fig:calib_params_energy_func_zetaem} for $v = \zetaem$. The increase in $b$ reflects the elongation rate. The increase in $m$ is small compared to its statistical uncertainty.
\par
Nevertheless, the $14\%$ systematic uncertainty in the FD energy scale~\cite{2020_Auger_sdSpectrum} must be propagated to the parameters and quantified, as it affects the inferred values of $\expval{v}$ required to reproduce shifts in $\expval{\xmax}$.
\par
\par
To first order, the impact of the energy-scale systematic uncertainty is estimated by fitting the energy dependence of $m$ and $b$ with linear functions of $\log_{10}(E_0/\mathrm{eV})$ and evaluating them at
\[
\log_{10} E_{\min} = 18.7 - \log_{10}(1.14), \quad
\log_{10} E_{\max} = 18.7 + \log_{10}(1.14).
\]
The resulting parameter sets, $(m_{E_{\min}}, b_{E_{\min}})$ and $(m_{E_{\max}}, b_{E_{\max}})$, are then used to compute $\expval{v_{E_{\min}}}$ and $\expval{v_{E_{\max}}}$, corresponding to $E_{\min}$ and $E_{\max}$, respectively, for a fixed value of $\expval{\xmax}$. The associated systematic uncertainty is defined as
\[
\sigma_{\mathrm{calib}}(v) = \frac{|\expval{v_{E_{\max}}} - \expval{v_{E_{\min}}}|}{2}.
\]

This contribution increases the total systematic uncertainty by at most $\sim 6\%$, across all hadronic interaction models and production variables. It is therefore subdominant compared to the uncertainty associated with the shift in $\expval{\xmax}$ and remains within the accuracy of the overall uncertainty estimation described in the main text. For this reason, it is neglected.

\subsection{Values of changed production variables of proton-air interactions}
Tab.~\ref{tab:shifted_primary_interaction_variables} lists the values of the proton--air multiparticle production variables that reproduce the $\expval{\xmax}$ shifts inferred from Auger data, together with the corresponding default predictions of the hadronic interaction models used in the main text.
\begin{center}
\begin{table}[ht!]
\caption{Default and shifted values of multiparticle production variables of first proton-air interactions (see Sec.~\ref{sec:first_interaction}) derived from the shifts in $\expval{\xmax}$~\cite{2024_Auger_Xmaxs1000fits}. The shifted values are identified by the prefix ``Data".
}\label{tab:shifted_primary_interaction_variables}
\begin{tabular}{l | c | c | c | c | c}
\hline \hline
Model & $\expval{\ln \elasticity / \multtotal}$ & $\expval{\ln \elasticity}$ & $\expval{\alphahad}$ & $\expval{\zetahad}$ & $\expval{\zetaem}$ \\ \hline \hline
\eposr{} & $-6.53$ & $-1.27$ & $0.81$ & $1.86$ & $0.64$ \\
\epos{}  & $-6.94$ & $-1.44$ & $0.79$ & $2.01$ & $0.70$ \\
\qIII{}  & $-7.11$ & $-1.56$ & $0.79$ & $2.15$ & $0.78$ \\
\qII{}   & $-7.31$ & $-1.69$ & $0.79$ & $2.25$ & $0.85$ \\
\sibe{}  & $-6.69$ & $-1.16$ & $0.81$ & $1.78$ & $0.61$ \\
\sibd{}  & $-6.73$ & $-1.15$ & $0.81$ & $1.77$ & $0.61$ \\
\hline \hline
Data$\,+\,$\epos{} & $-6.41 \pm 0.03_{-0.37}^{+0.32}$ & $-1.05 \pm 0.07_{-0.28}^{+0.30}$ & $0.821 \pm 0.004_{-0.017}^{+0.018}$ & $1.67 \pm 0.07_{-0.26}^{+0.27}$ & $0.55 \pm 0.03_{-0.13}^{+0.13}$ \\ 
Data$\,+\,$\qII{} & $-6.15 \pm 0.04_{-0.31}^{+0.28}$ & $-0.87 \pm 0.11_{-0.27}^{+0.30}$ & $0.832 \pm 0.006_{-0.017}^{+0.018}$ & $1.50 \pm 0.10_{-0.24}^{+0.27}$ & $0.47 \pm 0.04_{-0.12}^{+0.12}$ \\ 
Data$\,+\,$\sibd{} & $-5.99 \pm 0.05_{-0.36}^{+0.31}$ & $-0.76 \pm 0.13_{-0.31}^{+0.34}$ & $0.838 \pm 0.008_{-0.019}^{+0.020}$ & $1.41 \pm 0.11_{-0.28}^{+0.30}$ & $0.42 \pm 0.05_{-0.14}^{+0.14}$ \\ 
\hline \hline
\end{tabular}
\end{table}
\end{center}

\subsection{Interpretation of \boldmath{$\expval{\zetahad}$} and \boldmath{$\expval{\zetaem}$}}

\begin{figure}[ht!]
    \centering
    \includegraphics[width=6.5in]{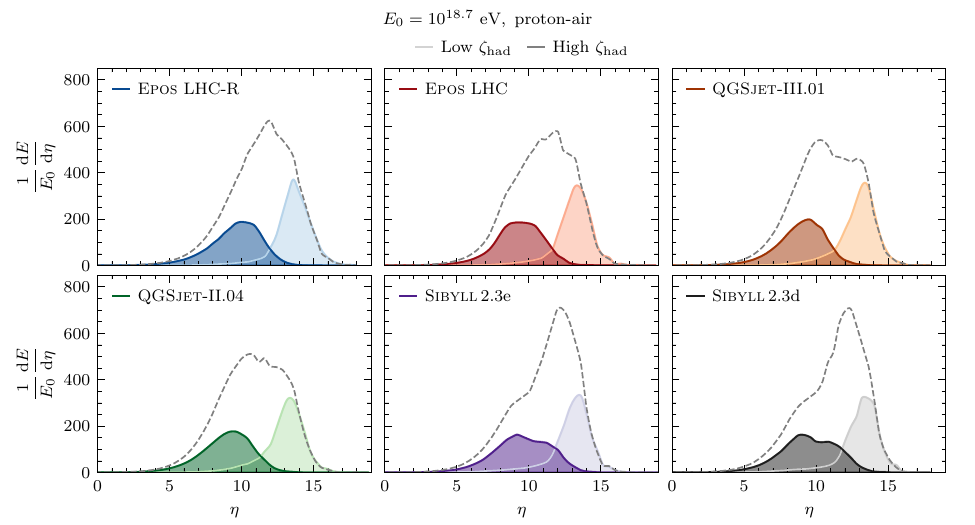}
    \caption{Differential energy flow into to the hadronic sector of proton-air interactions at $E_0 = 10^{18.7}$ eV in the first (light shaded areas) and last (dark shaded areas) quartiles of $\zetahad$, for the hadronic interaction models considered in the main text. The overall differential energy flow is represented by the dashed grey envelop.}
    \label{fig:eflow_per_zetahad}
\end{figure}

The relationship between $\zetahad$, $\zetaem$, and hadron production across different regions of phase space is illustrated by the $\zeta$ dependence of the differential energy flow in proton--air interactions at $E_0 = 10^{18.7}$ eV. Fig.~\ref{fig:eflow_per_zetahad} shows this dependence by splitting the ensemble of proton--air interactions into two subsets: the lowest-$\zetahad$ quartile, containing the 25 \% of events with the lowest $\zetahad$ values, and the highest-$\zetahad$ quartile, containing the $25\%$ of events with the highest values of $\zetahad$. The procedure was repeated for the main hadronic interaction models used in the analysis presented in the main text. The underlying grey distributions correspond to the energy flow integrated over all interactions, irrespective of their $\zetahad$ values.
\par
\par
For all hadronic interaction models, the energy flow in interactions with low $\zetahad$ peaks sharply at higher pseudorapidities. These interactions produce high-energy secondaries in the far-forward phase-space region, together with very-low-energy particles, leading to a very asymmetric partition of the primary energy among secondary particles. In contrast, high values of $\zetahad$ correspond to broader energy flows peaking at lower pseudorapidities. This behavior is characteristic of interactions producing a large number of secondary particles, all carrying a small fraction of the primary energy. A similar shift of the energy-flow peak towards higher $\eta$ is observed as $\zetaem$ decreases. However, this peak becomes more suppressed because the decrease in $\zetaem$ is also driven by a decrease in the total electromagnetic energy, which corresponds to the integral of the energy flow and suppresses neutral-pion production.
\par
The connection between $\zetahad$, $\zetaem$, and the pseudorapidity $\eta$ is more explicit using the relations from~\cite{2025_miguel_xmaxmodel}
\begin{equation}
\begin{aligned}
    \zetahad & = - \sum_{i = 1}^{\multhad} x_i \eta_i + \alphahad \ln\left(\frac{2E_0}{Q} \right) \equiv - \etahad + \alphahad \ln\left(\frac{2E_0}{Q} \right) \\
    \zetaem & = - \sum_{i = 1}^{\multem} x_i \eta_i + (1 - \alphahad) \ln\left(\frac{2E_0}{Q} \right) \equiv - \etaem + (1 - \alphahad) \ln\left(\frac{2E_0}{Q} \right)
\end{aligned}
,
\end{equation}
\par
\begin{figure}[ht!]
    \centering
    \includegraphics[width=5.5in]{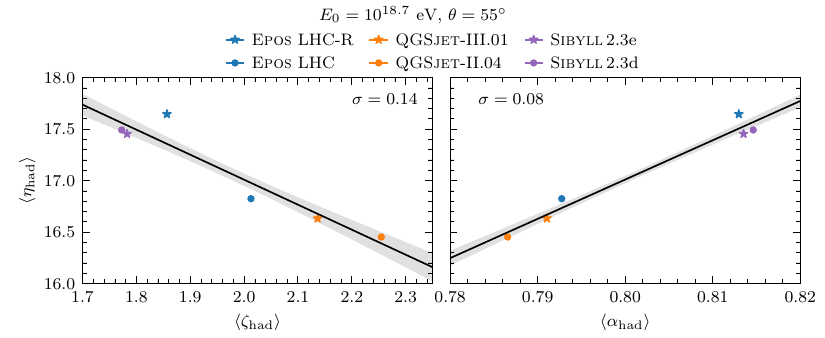}
    \caption{Inter-model calibrations of $\expval{\etahad}$ and $\expval{\zetahad}$ (left) or $\expval{\alphahad}$ (right). The variable $\etahad$ is the energy-weighted sum of pseudorapidities of hadronically interacting particles, in a given proton-air interaction. The standard deviation of residuals of the models about the curves is denoted by $\sigma$.}
    \label{fig:zetahad_etahad_calib}
\end{figure}

For the hadronic sector, decreasing $\expval{\zetahad}$ while increasing $\expval{\alphahad}$, as favored by Auger data, necessarily implies an increase in the average energy-weighted pseudorapidity, $\expval{\etahad}$. This behavior traces the drift of the energy flow across phase space discussed above. Moreover, Fig.~\ref{fig:zetahad_etahad_calib} shows that this evolution follows the same universal trend across several hadronic interaction models.
\par
\par
\begin{figure}[ht!]
    \centering
    \includegraphics[width=5.5in]{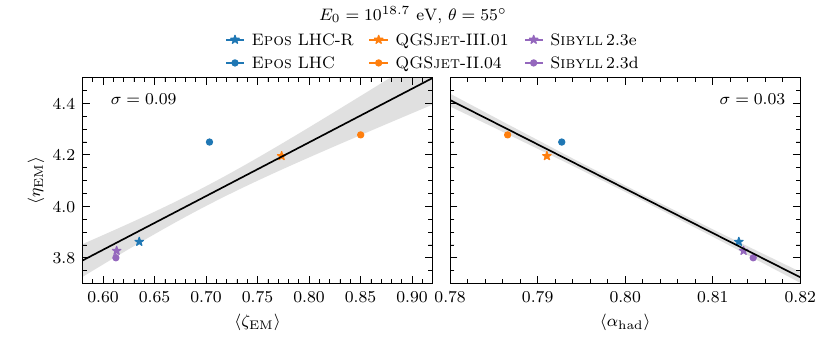}
    \caption{Inter-model relations between $\expval{\etaem}$ and $\expval{\zetaem}$ (left) or $\expval{\alphaem}$ (right). The variable $\etaem$ is the energy-weighted sum of pseudorapidities of secondaries of the EM sector (defined in Sec.~\ref{sec:first_interaction} of the main text), in a given proton-air interaction. The standard deviation of residuals of the models about the curves is denoted by $\sigma$.}
    \label{fig:zetaem_etaem_calib}
\end{figure}

Similar universal trends are observed for the relationships between $\expval{\etaem}$, $\expval{\zetaem}$, and $\expval{\alphahad}$, shown in Fig.~\ref{fig:zetaem_etaem_calib}, although with a larger dispersion than in the hadronic sector.
\par
\par
Nevertheless, the universality of the curves displayed in Figures~\ref{fig:zetahad_etahad_calib} and \ref{fig:zetaem_etaem_calib} enables the determination of $\expval{\etahad}$ and $\expval{\etaem}$ from kinematic variables directly accessible to accelerator experiments.

\subsection{Relation between increased {\boldmath $\alphahad$} and the Muon Puzzle}

\begin{figure}[ht!]
    \centering
    \includegraphics[width=3.1in]{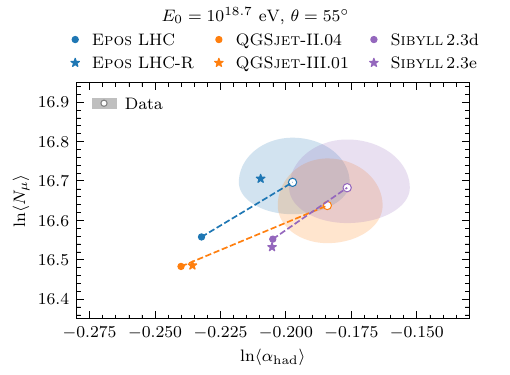}
    \caption{ $\ln \expval{\alphahad}$ vs $\ln \expval{N_\mu}$ for different high-energy hadronic interaction models. The solid markers correspond to the default values for each hadronic interaction model and the empty markers to the shifted values determined from Ref.~\cite{2024_Auger_Xmaxs1000fits} and this work. The slope of the dashed lines gives the amplification of the increase in $\ln \expval{\alphahad}$ required to account for the measured $\ln \expval{N_\mu}$.}
    \label{fig:alpha_nmu_calibration_together}
\end{figure}

To better compare the muon rescaling factors across hadronic interaction models and the corresponding values of $\expval{\alphahad}$ preferred by the shifted $\expval{\xmax}$ values, we merge the panels of Fig.~\ref{fig:alpha_nmu_calibration_per_model} into a single panel, shown in Fig.~\ref{fig:alpha_nmu_calibration_together}.
\par
The uncertainties on the amplification factors of $\expval{\alphahad}$ required to match the values of $R_{\mu}$ quoted in the main text, $g_{\text{mod}}$, are obtained from the slopes of lines tangent to the uncertainty ellipses in $(\ln \langle \alphahad\rangle, \ln \langle N_\mu\rangle)$ space, evaluated at the modified values and passing through the default hadronic interaction model predictions.

\subsection{Experimental values of the muon rescaling {\boldmath$R_\mu$}}
The values of $R_\mu$ used in this work are derived from the hadronic rescaling factor
$R_{\rm had}$ reported in Ref.~\cite{2024_Auger_Xmaxs1000fits}, following the prescription of Ref.~\cite{2019_Cazon_UHECR18_HadInts} and assuming
electromagnetic scaling $R_E = 1$. The factor $R_{\rm had}$ is extracted from the shower
signal at 1000~m from the shower core, $S(1000)$, which is sensitive to the muon content
of the shower. Using simulations of proton-induced air showers with
$E_0 = 10^{19}\,\mathrm{eV}$ and $\theta = 60^\circ$, we verified that the relative
differences in the muon signal inferred from $S(1000)$ match those of the total
number of muons at the ground, $N_\mu$, to approximately 2\%, in agreement with Ref.~\cite{2026_Blazek_mochi}. The difference between using the
signal at 1000~m and accounting for all core distances is therefore negligible for the
purposes of the present analysis, and the inferred $R_\mu$ values can be applied directly
as a rescaling of $N_\mu$ relative to the reference model.

\end{document}

%% file: revtex_authorlist.tex

\affiliation{Centro At\'omico Bariloche and Instituto Balseiro (CNEA-UNCuyo-CONICET), San Carlos de Bariloche, Argentina}
\affiliation{Departamento de F\'\i{}sica and Departamento de Ciencias de la Atm\'osfera y los Oc\'eanos, FCEyN, Universidad de Buenos Aires and CONICET, Buenos Aires, Argentina}
\affiliation{IFLP, Universidad Nacional de La Plata and CONICET, La Plata, Argentina}
\affiliation{Instituto de Astronom\'\i{}a y F\'\i{}sica del Espacio (IAFE, CONICET-UBA), Buenos Aires, Argentina}
\affiliation{Instituto de F\'\i{}sica de Rosario (IFIR) -- CONICET/U.N.R.\ and Facultad de Ciencias Bioqu\'\i{}micas y Farmac\'euticas U.N.R., Rosario, Argentina}
\affiliation{Instituto de Tecnolog\'\i{}as en Detecci\'on y Astropart\'\i{}culas (CNEA, CONICET, UNSAM), and Universidad Tecnol\'ogica Nacional -- Facultad Regional Mendoza (CONICET/CNEA), Mendoza, Argentina}
\affiliation{Instituto de Tecnolog\'\i{}as en Detecci\'on y Astropart\'\i{}culas (CNEA, CONICET, UNSAM), Buenos Aires, Argentina}
\affiliation{International Center of Advanced Studies and Instituto de Ciencias F\'\i{}sicas, ECyT-UNSAM and CONICET, Campus Miguelete -- San Mart\'\i{}n, Buenos Aires, Argentina}
\affiliation{Laboratorio Atm\'osfera -- Departamento de Investigaciones en L\'aseres y sus Aplicaciones -- UNIDEF (CITEDEF-CONICET), Argentina}
\affiliation{Observatorio Pierre Auger, Malarg\"ue, Argentina}
\affiliation{Observatorio Pierre Auger and Comisi\'on Nacional de Energ\'\i{}a At\'omica, Malarg\"ue, Argentina}
\affiliation{Universidad Tecnol\'ogica Nacional -- Facultad Regional Buenos Aires, Buenos Aires, Argentina}
\affiliation{Adelaide University, Adelaide, S.A., Australia}
\affiliation{Universit\'e Libre de Bruxelles (ULB), Brussels, Belgium}
\affiliation{Vrije Universiteit Brussels, Brussels, Belgium}
\affiliation{Centro Brasileiro de Pesquisas Fisicas, Rio de Janeiro, RJ, Brazil}
\affiliation{Centro Federal de Educa\c{c}\~ao Tecnol\'ogica Celso Suckow da Fonseca, Petropolis, Brazil}
\affiliation{Universidade de S\~ao Paulo, Escola de Engenharia de Lorena, Lorena, SP, Brazil}
\affiliation{Universidade de S\~ao Paulo, Instituto de F\'\i{}sica de S\~ao Carlos, S\~ao Carlos, SP, Brazil}
\affiliation{Universidade de S\~ao Paulo, Instituto de F\'\i{}sica, S\~ao Paulo, SP, Brazil}
\affiliation{Universidade Estadual de Campinas (UNICAMP), IFGW, Campinas, SP, Brazil}
\affiliation{Universidade Estadual de Feira de Santana, Feira de Santana, Brazil}
\affiliation{Universidade Federal do ABC, Santo Andr\'e, SP, Brazil}
\affiliation{Universidade Federal do Paran\'a, Setor Palotina, Palotina, Brazil}
\affiliation{Universidade Federal do Rio de Janeiro, Instituto de F\'\i{}sica, Rio de Janeiro, RJ, Brazil}
\affiliation{Universidad de Medell\'\i{}n, Medell\'\i{}n, Colombia}
\affiliation{Universidad Industrial de Santander, Bucaramanga, Colombia}
\affiliation{Charles University, Faculty of Mathematics and Physics, Institute of Particle and Nuclear Physics, Prague, Czech Republic}
\affiliation{Institute of Physics of the Czech Academy of Sciences, Prague, Czech Republic}
\affiliation{Palacky University, Olomouc, Czech Republic}
\affiliation{CNRS/IN2P3, IJCLab, Universit\'e Paris-Saclay, Orsay, France}
\affiliation{Laboratoire de Physique Nucl\'eaire et de Hautes Energies (LPNHE), Sorbonne Universit\'e, Universit\'e de Paris, CNRS-IN2P3, Paris, France}
\affiliation{Universit\'e Paris-Saclay, CNRS/IN2P3, IJCLab, Orsay, France}
\affiliation{Bergische Universit\"at Wuppertal, Department of Physics, Wuppertal, Germany}
\affiliation{Karlsruhe Institute of Technology (KIT), Institute for Experimental Particle Physics, Karlsruhe, Germany}
\affiliation{Karlsruhe Institute of Technology (KIT), Institut f\"ur Prozessdatenverarbeitung und Elektronik, Karlsruhe, Germany}
\affiliation{Karlsruhe Institute of Technology (KIT), Institute for Astroparticle Physics, Karlsruhe, Germany}
\affiliation{RWTH Aachen University, III.\ Physikalisches Institut A, Aachen, Germany}
\affiliation{TU Dortmund University, Department of Physics, Dortmund, Germany}
\affiliation{Universit\"at Hamburg, II.\ Institut f\"ur Theoretische Physik, Hamburg, Germany}
\affiliation{Universit\"at Siegen, Department Physik -- Experimentelle Teilchenphysik, Siegen, Germany}
\affiliation{Gran Sasso Science Institute, L'Aquila, Italy}
\affiliation{INFN Laboratori Nazionali del Gran Sasso, Assergi (L'Aquila), Italy}
\affiliation{INFN, Sezione di Catania, Catania, Italy}
\affiliation{INFN, Sezione di Lecce, Lecce, Italy}
\affiliation{INFN, Sezione di Milano, Milano, Italy}
\affiliation{INFN, Sezione di Napoli, Napoli, Italy}
\affiliation{INFN, Sezione di Roma ``Tor Vergata'', Roma, Italy}
\affiliation{INFN, Sezione di Torino, Torino, Italy}
\affiliation{Osservatorio Astrofisico di Torino (INAF), Torino, Italy}
\affiliation{Politecnico di Milano, Dipartimento di Scienze e Tecnologie Aerospaziali , Milano, Italy}
\affiliation{Universit\`a del Salento, Dipartimento di Matematica e Fisica ``E.\ De Giorgi'', Lecce, Italy}
\affiliation{Universit\`a dell'Aquila, Dipartimento di Scienze Fisiche e Chimiche, L'Aquila, Italy}
\affiliation{Universit\`a di Catania, Dipartimento di Fisica e Astronomia ``Ettore Majorana``, Catania, Italy}
\affiliation{Universit\`a di Milano, Dipartimento di Fisica, Milano, Italy}
\affiliation{Universit\`a di Napoli ``Federico II'', Dipartimento di Fisica ``Ettore Pancini'', Napoli, Italy}
\affiliation{Universit\`a di Palermo, Dipartimento di Fisica e Chimica ''E.\ Segr\`e'', Palermo, Italy}
\affiliation{Universit\`a di Roma ``Tor Vergata'', Dipartimento di Fisica, Roma, Italy}
\affiliation{Universit\`a Torino, Dipartimento di Fisica, Torino, Italy}
\affiliation{Benem\'erita Universidad Aut\'onoma de Puebla, Puebla, M\'exico}
\affiliation{Unidad Profesional Interdisciplinaria en Ingenier\'\i{}a y Tecnolog\'\i{}as Avanzadas del Instituto Polit\'ecnico Nacional (UPIITA-IPN), M\'exico, D.F., M\'exico}
\affiliation{Universidad Aut\'onoma de Chiapas, Tuxtla Guti\'errez, Chiapas, M\'exico}
\affiliation{Universidad Michoacana de San Nicol\'as de Hidalgo, Morelia, Michoac\'an, M\'exico}
\affiliation{Universidad Nacional Aut\'onoma de M\'exico, M\'exico, D.F., M\'exico}
\affiliation{Institute of Nuclear Physics PAN, Krakow, Poland}
\affiliation{University of \L{}\'od\'z, Faculty of High-Energy Astrophysics,\L{}\'od\'z, Poland}
\affiliation{Laborat\'orio de Instrumenta\c{c}\~ao e F\'\i{}sica Experimental de Part\'\i{}culas -- LIP and Instituto Superior T\'ecnico -- IST, Universidade de Lisboa -- UL, Lisboa, Portugal}
\affiliation{``Horia Hulubei'' National Institute for Physics and Nuclear Engineering, Bucharest-Magurele, Romania}
\affiliation{Institute of Space Science, Bucharest-Magurele, Romania}
\affiliation{Center for Astrophysics and Cosmology (CAC), University of Nova Gorica, Nova Gorica, Slovenia}
\affiliation{Experimental Particle Physics Department, J.\ Stefan Institute, Ljubljana, Slovenia}
\affiliation{Instituto Galego de F\'\i{}sica de Altas Enerx\'\i{}as (IGFAE), Universidade de Santiago de Compostela, Santiago de Compostela, Spain}
\affiliation{IMAPP, Radboud University Nijmegen, Nijmegen, The Netherlands}
\affiliation{Nationaal Instituut voor Kernfysica en Hoge Energie Fysica (NIKHEF), Science Park, Amsterdam, The Netherlands}
\affiliation{Stichting Astronomisch Onderzoek in Nederland (ASTRON), Dwingeloo, The Netherlands}
\affiliation{Case Western Reserve University, Cleveland, OH, USA}
\affiliation{Colorado School of Mines, Golden, CO, USA}
\affiliation{Department of Physics and Astronomy, Lehman College, City University of New York, Bronx, NY, USA}
\affiliation{Michigan Technological University, Houghton, MI, USA}
\affiliation{New York University, New York, NY, USA}
\affiliation{University of Chicago, Enrico Fermi Institute, Chicago, IL, USA}
\affiliation{University of Delaware, Department of Physics and Astronomy, Bartol Research Institute, Newark, DE, USA}

\author{A.~\surname{Abdul Halim}}
\affiliation{Adelaide University, Adelaide, S.A., Australia}

\author{P.~\surname{Abreu}}
\affiliation{Laborat\'orio de Instrumenta\c{c}\~ao e F\'\i{}sica Experimental de Part\'\i{}culas -- LIP and Instituto Superior T\'ecnico -- IST, Universidade de Lisboa -- UL, Lisboa, Portugal}

\author{M.~\surname{Aglietta}}
\affiliation{Osservatorio Astrofisico di Torino (INAF), Torino, Italy}
\affiliation{INFN, Sezione di Torino, Torino, Italy}

\author{M.~\surname{Ahmed}}
\affiliation{CNRS/IN2P3, IJCLab, Universit\'e Paris-Saclay, Orsay, France}

\author{I.~\surname{Allekotte}}
\affiliation{Centro At\'omico Bariloche and Instituto Balseiro (CNEA-UNCuyo-CONICET), San Carlos de Bariloche, Argentina}

\author{K.~\surname{Almeida Cheminant}}
\affiliation{Nationaal Instituut voor Kernfysica en Hoge Energie Fysica (NIKHEF), Science Park, Amsterdam, The Netherlands}
\affiliation{IMAPP, Radboud University Nijmegen, Nijmegen, The Netherlands}

\author{R.~\surname{Aloisio}}
\affiliation{Gran Sasso Science Institute, L'Aquila, Italy}
\affiliation{INFN Laboratori Nazionali del Gran Sasso, Assergi (L'Aquila), Italy}

\author{J.~\surname{Alvarez-Mu\~niz}}
\affiliation{Instituto Galego de F\'\i{}sica de Altas Enerx\'\i{}as (IGFAE), Universidade de Santiago de Compostela, Santiago de Compostela, Spain}

\author{A.~\surname{Ambrosone}}
\affiliation{Gran Sasso Science Institute, L'Aquila, Italy}
\affiliation{INFN Laboratori Nazionali del Gran Sasso, Assergi (L'Aquila), Italy}

\author{J.~\surname{Ammerman Yebra}}
\affiliation{Instituto Galego de F\'\i{}sica de Altas Enerx\'\i{}as (IGFAE), Universidade de Santiago de Compostela, Santiago de Compostela, Spain}

% ...

\collaboration{The Pierre Auger Collaboration}
\email{spokespersons@auger.org}
\homepage{http://www.auger.org}
\noaffiliation

%% file: acknowledgments.tex
\section*{Acknowledgments}

\begin{sloppypar}
The successful installation, commissioning, and operation of the Pierre
Auger Observatory would not have been possible without the strong
commitment and effort from the technical and administrative staff in
Malarg\"ue. We are very grateful to the following agencies and
organizations for financial support:
\end{sloppypar}

\begin{sloppypar}
Argentina -- Comisi\'on Nacional de Energ\'\i{}a At\'omica; Agencia Nacional de
Promoci\'on Cient\'\i{}fica y Tecnol\'ogica (ANPCyT); Consejo Nacional de
Investigaciones Cient\'\i{}ficas y T\'ecnicas (CONICET); Gobierno de la
Provincia de Mendoza; Municipalidad de Malarg\"ue; NDM Holdings and Valle
Las Le\~nas; in gratitude for their continuing cooperation over land
access; Australia -- the Australian Research Council; Belgium -- Fonds
de la Recherche Scientifique (FNRS); Research Foundation Flanders (FWO),
Marie Curie Action of the European Union Grant No.~101107047; Brazil --
Minist\'erio da Ci\^encia, Tecnologia e Inova\c{c}\~ao (MCTI); Czech Republic --
GACR 24-13049S, CAS LQ100102401, MEYS LM2023032,
CZ.02.1.01/0.0/0.0/16{\textunderscore}013/0001402, CZ.02.1.01/0.0/0.0/18{\textunderscore}046/0016010
and CZ.02.1.01/0.0/0.0/17{\textunderscore}049/0008422 and
CZ.02.01.01/00/22{\textunderscore}008/0004632; France -- Centre de Calcul IN2P3/CNRS;
Centre National de la Recherche Scientifique (CNRS); Institut National
de Physique Nucl\'eaire et de Physique des Particules (IN2P3/CNRS);
Germany -- Bundesministerium f\"ur Forschung, Technologie und Raumfahrt
(BMFTR); Deutsche Forschungsgemeinschaft (DFG); Ministerium f\"ur Finanzen
Baden-W\"urttemberg; Helmholtz Alliance for Astroparticle Physics (HAP);
Hermann von Helmholtz-Gemeinschaft Deutscher Forschungszentren e.V.;
Ministerium f\"ur Kultur und Wissenschaft des Landes Nordrhein-Westfalen;
Ministerium f\"ur Wissenschaft, Forschung und Kunst des Landes
Baden-W\"urttemberg; Italy -- Istituto Nazionale di Fisica Nucleare
(INFN); Istituto Nazionale di Astrofisica (INAF); Ministero
dell'Universit\`a e della Ricerca (MUR); CETEMPS Center of Excellence;
Ministero degli Affari Esteri (MAE), ICSC Centro Nazionale di Ricerca in
High Performance Computing, Big Data and Quantum Computing, funded by
European Union NextGenerationEU, reference code CN{\textunderscore}00000013; M\'exico --
Consejo Nacional de Ciencia y Tecnolog\'\i{}a (CONACYT-SECHTI)
No.~CB-A1-S-46703, Universidad Nacional Aut\'onoma de M\'exico (UNAM)
PAPIIT-IN114924; Benem\'erita Universidad Aut\'onoma de Puebla (BUAP), VIEP
and Laboratorio Nacional de Superc\'omputo del Sureste de M\'exico (LNS);
and Benem\'erita Universidad Aut\'onoma de Chiapas (UNACH); The Netherlands
-- Ministry of Education, Culture and Science; Netherlands Organisation
for Scientific Research (NWO); Dutch national e-infrastructure with the
support of SURF Cooperative; Poland -- Ministry of Science and Higher
Education, grant No.~2026/WK/05; National Science Centre, grant
No.~2022/45/B/ST9/02163; Portugal -- Portuguese national funds and FEDER
funds within Programa Operacional Factores de Competitividade through
Funda\c{c}\~ao para a Ci\^encia e a Tecnologia (COMPETE); Romania -- Ministry of
Education and Research, contract no.~30N/2023 under Romanian National
Core Program LAPLAS VII, and grant no.~PN 23 21 01 02; Slovenia --
Slovenian Research and Innovation Agency, grants P1-0031, I0-0033; Spain
-- Ministerio de Ciencia, Innovaci\'on y Universidades/Agencia Estatal de
Investigaci\'on MICIU/AEI /10.13039/501100011033 (PID2022-140510NB-I00,
PCI2023-145952-2, CNS2024-154676, and Mar\'\i{}a de Maeztu CEX2023-001318-M),
Xunta de Galicia (CIGUS Network of Research Centers, Consolidaci\'on
ED431C-2025/11 and ED431F-2022/15) and European Union ERDF; USA --
Department of Energy, Contracts No.~DE-AC02-07CH11359,
No.~DE-FR02-04ER41300, No.~DE-FG02-99ER41107 and No.~DE-SC0011689;
National Science Foundation, Grant No.~0450696, and NSF-2013199; The
Grainger Foundation; Astrophysics Centre for Multi-messenger studies in
Europe (ACME) EU Grant No 101131928; and UNESCO.
\end{sloppypar}